\documentclass{aa}
\usepackage{graphics}
\usepackage{times}
\def\kms{km\,s$^{-1}$}
\def\hb{H$\beta$}
\def\ha{H$\alpha$}
\def\nii{\lbrack\ion{N}{ii}\rbrack}
\def\oiii{\lbrack\ion{O}{iii}\rbrack}

\def\sii{\lbrack\ion{S}{ii}\rbrack}

\def\oi{\lbrack\ion{O}{i}\rbrack}
\def\roiii{$\lambda$5007$/$H$\beta$}
\def\rheii{$\lambda$4686$/$H$\beta$}
\def\rnii{$\lambda$6584$/$H$\alpha$}
\def\roi{$\lambda$6300$/$H$\alpha$}

\begin{document}

\thesaurus{11                
              (11.01.2;      
               11.14.1;      
               11.19.1;      
               11.09.1)}     
\title{NGC 5252 -- a Liner undercover
\thanks{Based on observations collected at the Observatoire de Haute-Provence
(CNRS, France) and {\it{Hubble Space Telescope}} data obtained from the Space 
Telescope European Coordinating Facility (ST-ECF) Archive.}
}
\author{A.C.\,Gon\c{c}alves \and P.\,V\'eron \and M.-P.\,V\'eron-Cetty}
\offprints{A.C.\,Gon\c{c}alves, {\small{anabela@obs-hp.fr}} }
\institute{Observatoire de Haute-Provence (CNRS), F-04870 Saint Michel
l'Observatoire, France}
\date{Received 8 December 1997~/~Accepted 4 February 1998}
\maketitle


\begin{abstract}

Ground based long slit spectroscopic observations of the nuclear region of 
\object{NGC 5252}, and {\it{Hubble Space Telescope (HST)}} Faint Object 
Spectrograph (FOS) spectra of the nucleus and two bright knots located 
0\farcs36 NE and 0\farcs31 SW from it, show that the nuclear region 
exhibits the characteristics of a Liner with exceptionally strong 
\oi\ emission, all lines being broad (FWHM $\sim$ 1\,100 \kms), 
while the gas outside the nucleus has a typical Seyfert 2 spectrum with 
relatively narrow lines ($\sim$ 200--300 \kms). We suggest that all the 
emitting gas is photoionized by the hidden non-thermal nuclear source detected 
through near-infrared (Kotilainen \& Prieto 1995) and X-ray (Cappi et al. 
1996) observations, the ionizing continuum, in the case of the central Liner, 
being ``filtered'' by a matter-bounded highly ionized cloud, hidden from our 
view by the same material obscuring the central continuum source. \\


\end{abstract}


\section{Introduction}

NGC 5252 (1335+04) is a S0 galaxy at a redshift of z $\sim$ 0.023. 
Nuclear spectra have shown it to be a Seyfert 2 galaxy (V\'eron-Cetty \& 
V\'eron 1986; Huchra \& Burg 1992). However, Osterbrock \& Martel (1993) 
have found a weak broad \ha\ component in the nuclear region; Acosta-Pulido 
et al. (1996) have confirmed the presence of this broad component, with a 
measured FWHM of 2\,485 $\pm$ 78 \kms. Using a 3\arcsec-diameter aperture, 
Ruiz et al. (1994) detected the \ion{He}{i}\,$\lambda$1.083 $\mu$m emission 
line in the nucleus; in addition to a narrow component (527 \kms\ FWHM), 
this line shows a broad feature (1\,043 \kms\ FWHM), which was 
interpreted as the signature of a Seyfert~1 cloud. Goodrich et al. (1994) 
have reported a marginal detection of a broad Pa\,$\beta$ line.

Unger et al. (1987), having obtained a high resolution (0.75 \AA) slit 
spectrum of the nuclear region of NGC 5252, found the 
\oiii $\lambda\lambda$\,4959, 5007 lines to be double with a velocity 
separation of $\sim$ 180 \kms. Acosta-Pulido et al. (1996) obtained a 
2.0~\AA\ resolu\-tion spectrum centered on the nucleus; this spectrum, 
covering the red spectral region, was extracted on a length of 4\farcs6 
along the slit, which was 1\farcs0 wide. 
The \sii $\lambda\lambda$\,6716, 6731 and 
\ha +\nii $\lambda\lambda$\,6548, 6584 complexes were fitted with two 
Gaussians for each line, the velocity difference between the two components 
being $\sim$ 200 \kms; the need for an additional broad \ha\ component 
was already mentioned above.

\begin{table}[b]
\begin{center}
\caption{OHP and {\it{HST}} observing log}
\begin{flushleft}
\begin{tabular}{llccl}
\hline
 ID Label     &  Target   &  Date      & Exposure  &  \\
	      &           &            & Time (s)  &  \\
\hline
Y2YZ010BT     &  Nucleus  &  17.01.96  & 1\,000 & {\it{HST}} \\
Y2YZ010CT     &  NE knot  &  17.01.96  & 1\,190 & {\it{HST}} \\
Y2YZ010ET     &  SW knot  &  17.01.96  & 1\,600 & {\it{HST}} \\
NGC 5252~~\hb &   	  &  06.03.97  & 1\,200 & OHP \\
NGC 5252~~\ha &  	  &  09.03.97  & 1\,200 & OHP \\
\hline
\end{tabular}
\end{flushleft}
\end{center}
\label{obs_log}
\end{table}

Images taken in the light of the \oiii$\lambda$5007 line reveal a sharply 
defined biconical structure extending to a maximum radius of 48\arcsec, 
corresponding to 32 kpc, if H$\rm _{o}$ = 50 \kms\,Mpc$^{-1}$ (Tadhunter 
\& Tsvetanov 1989). Spectra (Durret \& Warin 1990) and images in the light 
of \oiii $\lambda$5007 and \ha+\nii\ (Haniff et al. 1991; Prieto \& 
Freudling 1996) show that the gas outside the nucleus has a very high 
\roiii\ ratio. {\it{HST}} narrow band images 
%
%
show that three bright knots dominate the line emission in the innermost 
1\arcsec; they are aligned along PA $\sim$ 35\degr\ with a total separation 
of $\sim$ 0\farcs7 and are embedded in fainter diffuse gas (Tsvetanov 
et al. 1996). 

\begin{table*}[t]
\begin{center}
\caption{Line profile fitting results for the OHP and {\it{HST}} spectra. 
Fluxes are in units of 10$^{-16}$\,erg\,cm$^{-2}$\,s$^{-1}$. FWHMs have been 
corrected for instrumental broadening}
\begin{flushleft}
\begin{tabular}{lrccrrrrrrr}
\hline
	          &  F(\ha)        &  \underline{\,$\lambda$6584\,}     & 
\underline{\,$\lambda$6300\,}      &  V\verb+   +     &  FWHM$\:$       &
                  &  F(\hb)        &  \underline{\,$\lambda$5007\,}     &
   V\verb+   +    &  FWHM$\:$     \\
                  &                &  \ha$\;$                           &
H$\alpha\;$       &  (\kms)	   &  (\kms)	                        &
		  &                &  \hb$\,\;\;$        		&
(\kms)	          &  (\kms)	  \\
\hline
OHP		  & 84\verb+ +     & 1.05			        &
0.26	                           & 141\verb+  +   & 165\verb+  +      &
		  & 57\verb+ +     & 7.04\verb+ +		        &
96\verb+  +       & 200\verb+  +  \\
	          & 231\verb+ +    & 0.90	                        &
0.29		  		   & -116\verb+  +  & 265\verb+  +      &
		  & 66\verb+ +     & 8.38\verb+ +                       &   
-146\verb+  +     & 225\verb+  +  \\
		  & 153\verb+ +    & 0.90                               &
0.73	                           & -261\verb+  +  & 1\,460\verb+  +   &
	          & 66\verb+ +     & 3.59\verb+ +                       &
-134\verb+  +     & 1\,080\verb+  +  \\
 & & & & & & & & & \\	
{\it{HST}} (SW knot) & 19\verb+ +  & 1.03	                        & 
0.29			           & 114\verb+  +   & 230\verb+  +      &
	          & 5\verb+ +      & 7.09\verb+ +                       &
99\verb+  +       & 210\verb+  +  \\
{\it{HST}} (NE knot) & 40\verb+ +  & 0.88	                        & 
0.26			           & -240\verb+  +  & 340\verb+  +      &
	          & 7\verb+ +      & 11.11\verb+ +                      &
-235\verb+  +     & 305\verb+  +  \\
{\it{HST}} (Nucleus) & 50\verb+ +  & 1.44	                        & 
1.35			           & -124\verb+  +  & 530\verb+  +      &
	          & 21\verb+ +     & 3.58\verb+ +                       &
-182\verb+  +     & 580\verb+  +  \\
	          & 137\verb+ +    & 0.63                               & 
0.49			           & -427\verb+  +  & 1\,590\verb+  +   &
	          & 36\verb+ +     & 2.72\verb+ +                       &
-501\verb+  +     & 1\,760\verb+  + \\
 & & & & & & & & & \\	
OHP - {\it{HST}} (Nuc) & 101\verb+ +  & 1.01	                        & 
0.23			           & 132\verb+  +   & 195\verb+  +      &
	          & 68\verb+ +     & 6.73\verb+ +                       &
96\verb+  +       & 240\verb+  +  \\
	          & 189\verb+ +    & 0.82	                        & 
0.19			           & -124\verb+  +  & 225\verb+  +      &
	          & 66\verb+ +     & 8.39\verb+ +                       &
-151\verb+  +     & 245\verb+  +  \\
\hline
\end{tabular}
\end{flushleft}
\end{center}
\end{table*}

J, H and K imaging with a seeing of 1\farcs5--1\farcs7 suggests the presence, 
in the nucleus, of a heavily reddened ($A\rm_{V} \sim$ 6 mag) non stellar 
source (Kotilainen \& Prieto 1995). ASCA observations show that NGC 5252 
is a relatively strong X-ray source ($L\rm _{X\,(0.7-10 kev)}$ = 
2.6\,10$^{43}$\,erg\,s$^{-1}$). A description of the spectrum with a single 
power law is ruled out; there is evidence for a strong soft excess. The 
best-fit partial covering model results in a flat ($\Gamma \sim$ 1.45 $\pm$ 
0.2) power-law continuum emitted by a source almost completely covered (at 
$\sim$ 94--97\%) by neutral matter ($N\rm_{H} \sim$ 
4.3\,10$^{22}$\,cm$^{-2}$) (Cappi et al. 1996). For galactic X-ray sources, 
the hydrogen column density $N\rm_{H}$ and the visual extinction $A\rm_{V}$ 
follow the relation: $A\rm_{V} \sim$ 5\,10$^{-22}\,N\rm_{H}$ (Gorenstein 1975; 
Reina \& Tarenghi 1973). If this is valid for galactic nuclei, the observed 
X-ray extinction would imply a visual extinction of $\sim$ 20 mag. Although 
this is significantly larger than the value derived from near IR observations, 
it confirms that the nuclear source is heavily reddened. Spectropolarimetry 
allowed only a marginal detection of a broad \ha\ component 
(You\-ng et al. 1996).

The presence of a broad \ha\ component in a Seyfert galaxy in which the 
nucleus is heavily obscured was surprising and induced us to observe this 
object.


\section{Observations and data analysis}

\subsection{Observations}

NGC 5252 was observed on March 6 and 9, 1997 with the spectrograph CARELEC 
(Lema\^{\i}tre et al. 1989) attached to the Cassegrain focus of the 
Observatoire de Haute-Provence 1.93m telescope. The detector was a 
512$\times$512 pixel, 27$\times$27 $\mu$m Tektronic CCD. We used a 600 
l\,mm$^{-1}$ grating giving a dispersion of 66~\AA\,mm$^{-1}$. A Schott 
GG 435 filter was used in the red spectral range, 
$\lambda\lambda$\,6305--7215 \AA; the wavelength 
range covered in the blue was $\lambda\lambda$\,4825--5730 \AA.

The slit width was 2\farcs0, corres\-ponding to a projected slit width on
the detector of 50 $\mu$m or 1.9 pixel; the slit P.A. was 90\degr\ for 
the blue spectrum and 180\degr\ for the red one. In each case, the galaxy 
nucleus was centered on the slit and 3 columns of the CCD ($\sim$ 3\farcs2) 
were extracted. The seeing was $\sim$ 3\arcsec\ on both nights; the 
resolution, as measured on the night sky emission lines, was $\sim$ 3.5~\AA\ 
FWHM in the blue, and $\sim$ 3.0~\AA\ FWHM in the red regions. The spectra 
were flux calibrated using the standard star Feige 66 (Massey et al. 1988), 
also used to correct the observations for the atmospheric absorption. 

To supplement our own observations, we searched the {\it{HST}} archives for 
spectra of the central region of NGC 5252. Three FOS (description by Ford 
\& Hartig 1990) spectra were retrie\-ved, corresponding to the nucleus and 
the two knots located 0\farcs36 NE and 0\farcs31 SW from it (details are 
given in Table~\ref{obs_log}). All three {\it{HST}} spectra were obtained 
under the same setting conditions, with the G570H grating 
(4.37~\AA\,diode$^{-1}$) and the FOS/RED detector (a 512 diodes linear 
array), resulting in a spectral range of $\sim$ 
$\lambda\lambda$\,4570--6820 \AA; a single 0\farcs26-diameter circular 
aperture was used. The spectra were submitted to the usual processes of 
substepping and overscanning, resulting in a 2\,064 pixel coverage. Since 
each diode corresponds to 0\farcs31 in the dispersion direction, the 
resolution was estimated at about 3.7~\AA\ FWHM. The {\it{HST}} data were 
processed by the calibration pipe\-line {\it Calfos}, which includes 
flat-fielding, subtraction of the background and sky, and wavelength and 
flux calibrations. All ({\it{HST}} and OHP) spectra were deredshifted to 
rest wavelengths with $z$ = 0.023.

The observing log is given in Table~\ref{obs_log}. \\

\subsection{Data analysis}
                                                                       
\begin{figure*}[t]
\resizebox{12cm}{!}{\includegraphics{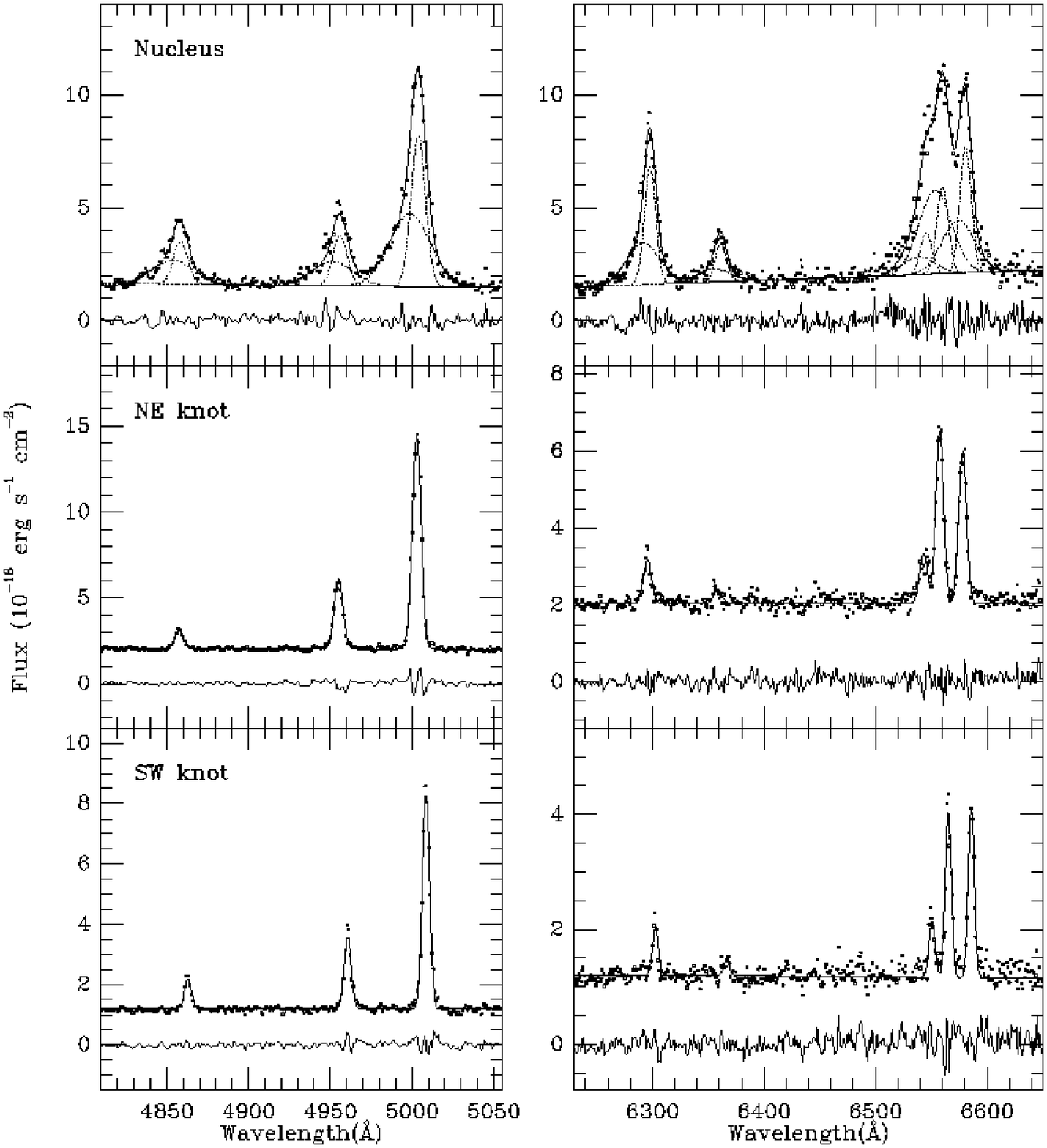}}
\hfill
\parbox[b]{55mm}{
\caption{Rest-wavelength {\it{HST}} spectra of the nucleus and knots, showing 
the spectral regions around \hb\ and \ha. The data points are represented as 
small squares and the best fit as a solid line; the lower line shows the 
residuals. On the upper panels (spectra of the nucleus), the individual 
components are also drawn. All spectra were shifted upwards, so the 
continuum and residuals do not overlap; the origin of the vertical (flux) 
scales are, therefore, arbitrary} 
\label{HST_obs}
}
\end{figure*}

The presence of an old star population with many strong absorption lines can 
make the line fitting analysis difficult, especially in the blue spectral 
region. Bica (1988) has shown this old star population to have similar spectra 
in all Morgan (1958, 1959) classes; therefore, a suitable fraction of the 
spectra of the elliptical galaxies NGC 5982 and NGC 4365 (used as templates 
for the blue and red spectra, respectively) was subtracted from the 
observations to remove the old stellar population contribution. NGC 5982 was 
observed on March 6, 1997 with the same instrumental setting used for 
NGC 5252, while NGC 4365 was observed on February 28, 1984 with the Boller 
\& Chivens spectrograph and the Image Dissector Scanner attached to the 
Cassegrain focus of the ESO 3.6 m telescope at La Silla; the dispersion was 
59~\AA\,mm$^{-1}$ and the resolution, 4.5~\AA\ FWHM. The subtraction of these 
template spectra from our observations resulted in much smoother continua and 
corrected the flux of the Balmer emission lines for the underlying Balmer 
absorption.

Inspection of our two-dimensional spectra shows the lines to be double and 
spatially extended in the nuclear region, confirming earlier results. 
{\it{HST}} and OHP spectra were analysed in terms of Gaussian components, 
as described in V\'eron et al. (1997). The emission lines \ha, 
\nii $\lambda\lambda$\,6548, 6584, \sii $\lambda\lambda$ 6716, 6731 
and \oi $\lambda\lambda$\,6300, 6363 (or \hb\ and \oiii $\lambda\lambda$\,4959, 
5007) were fitted by one or several sets of seven (three) Gaussian components; 
the width and redshift of each component in a set were taken to be the same. 
Therefore, in addition to the line intensities, the free parameters for each 
set of lines are one width and one redshift. The intensity ratios of the 
\nii $\lambda\lambda$\,6548, 6584, \oiii 
$\lambda\lambda$\,4959, 5007 and \oi $\lambda\lambda$\,6300, 6363 li\-nes 
were taken to be equal to 3.00, 2.96 and 3.11, respectively (Osterbrock 1974).

Fitting our large aperture red spectrum with two sets of Gaussians gives 
unsatisfactory results, with large residuals not only for the \ha+\nii\ 
complex, but also for the \oi\ lines, which have an obvious blue wing. 
The best solution is not obtained by adding a broad \ha\ component, but rather 
a third set of Gaussians. This third set of components has a relatively broad 
width ($\sim$ 1\,460 \kms), \rnii\ = 0.90 and extremely strong \oi\ lines 
($\lambda$6300$/$H$\alpha$ = 0.73). We obtained only an upper limit to the 
strength of the broad component of the \sii\ lines; while the total 
\sii\ flux relative to \ha\ for the two narrow components is 1.2 and 0.9 
respectively, this ratio is $<$ 0.5 for the broad component. The same model, 
with three sets of components, also succeeds in matching the blue spectrum, 
one set having a large width ($\sim$ 1\,080 \kms) and \roiii\ = 3.59. The 
width found for the \ha\ component is much larger than that of the equivalent 
\hb\ line; this, however, may not be significant since the errors in the 
\ha\ width should be larger, the narrow \ha\ and \nii\ components having a 
considerable relative strength. This ``broad'' line component has the 
characteristics of a Liner, while the two ``narrow'' line components are 
Seyfert~2-like, with weak \oi\ lines and strong \oiii\ emission (see 
Table~2). Although the width of the ``broad'' component may seem 
large for a Seyfert~2 or a Liner, it is not exceptional as the line width of 
the prototype Seyfert 2 galaxy NGC 1068 is $\sim$ 1\,500 \kms\ 
(Marconi et al. 1996). 

\begin{figure*}[t]
\resizebox{12cm}{!}{\includegraphics{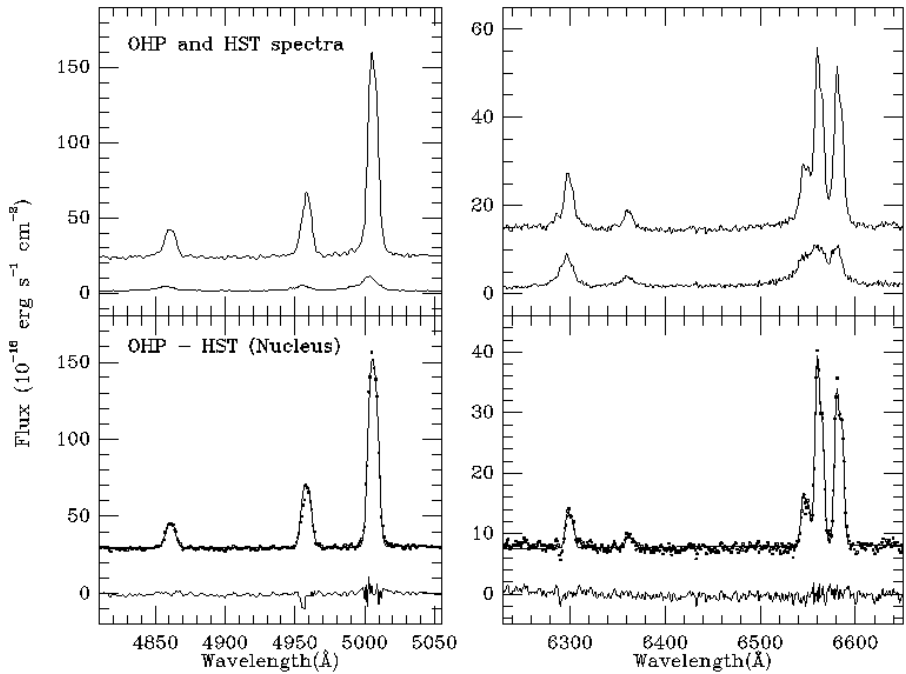}}
\hfill
\parbox[b]{55mm}{
\caption{The upper panels show the larger aperture OHP spectra (upper curves) 
and the {\it{HST}} nuclear spectra (lower curves), at rest-wavelengths. The 
lower panels show the difference of the two spectra (small squares), the best 
fit with two sets of narrow components (solid line) and the residuals (lower 
solid line). The spectra in the lower panel were shifted upwards by an 
arbitrary amount for clarity}
\label{OHP_obs}
}
\end{figure*}

We found no evidence for the presence of broad Balmer components typical 
of Seyfert 1 nuclei. The broad \ha\ component observed by Osterbrock \& 
Martel (1993) and Acosta-Pulido et al. (1996) does not really seem to exist; 
this feature is rather due to the unresolved blend of the \ha\ and 
\nii\ components. Whittle (1985) has mentioned the possibility of 
misidentifying weak relatively broad wings to the \ha\ and \nii\ li\-nes with 
a broad ($\sim$ 2\,000 \kms) \ha\ component; we have an illustration of such 
a possibility in IRAS 13197$-$1627: Aguero et al. (1994) have fitted \hb\ and 
the \ha+\nii\ complex with a set of narrow Gaussian components, adding a 
bro\-ad component to the Balmer lines; however, Young et al. (1996) showed 
that the \ha+\nii\ complex can be very satisfactorily fitted by two sets of 
Gaussians (with FWHMs of 400 and 1\,350 \kms\ and a \rnii\ ratio of 1.94 and 
1.35, respectively) without adding a broad \ha\ component. 

Fitting the {\it{HST}} spectra showed that both the NE and SW knots 
(Fig.~\ref{HST_obs}) have Seyfert~2-like spectra with relatively narrow 
lines ($\sim$ 325 and 220 \kms\ FWHM res\-pectively, corrected for the 
instrumental broadening) and a velocity difference of $\sim$ 345 
\kms\ (the SW knot being redshifted with respect to the NE knot). 
The observed line ratios are \roiii\ = 11.11 (7.09), \roi\ = 0.26 (0.29) and 
$\lambda$6584 $/$ H$\alpha$ = 0.88 (1.03) for the NE (SW) knot, respectively. 
The nucleus spectrum (Fig.~\ref{HST_obs}) is quite different: the lines are 
broad and have a complex profile; they have been fitted with two sets of 
components and have line ratios typical of Liners (see Table~2).

Our larger aperture (2\arcsec $\times$ 3\farcs2) included both the nucleus 
and the two bright knots (Fig.~\ref{OHP_obs}, upper panels). Subtracting 
the {\it{HST}} nucleus spectrum from our own, without any scaling, resulted 
in a spectrum which is well fitted by a set of two narrow components 
(Fig.~\ref{OHP_obs}, lower panels) showing that the broad lines come 
exclusively from the small 0\farcs26 aperture centered on the nucleus. 
The velocity difference between the two narrow line systems in the resulting 
spectrum is $\sim$ 250 \kms, the FWHM is approximately the same in both 
systems ($\sim$ 225 \kms) and the line ratios (\roiii\ = 8.39 (6.73), 
\roi\ = 0.19 (0.23) and \rnii\ = 0.82 (1.01) for the blueshifted and 
redshifted components, respectively) are very similar to those measured on 
the bright knots. The sum of the \ha\ fluxes of the two knots is equal to 
$\sim$ 20\% of the total \ha\ flux measured on the differential 
(OHP $-$ {\it{HST}}\,) spectrum, confirming the finding of Tsvetanov et 
al. (1996), that the knots are embedded in faint diffuse gas. 


\section{Discussion and Conclusions}
                        
The ionization mechanism within Liners is still a subject of debate, mainly 
because their emission line spectrum can be reasonably reproduced by very 
different models, based on shock excitation, hot stars, or non-stellar 
photoionization. It has been shown that Liners and Seyfert 2 galaxies could 
be photoionized by the same non-thermal continuum, with a lower ionization 
parameter for Liners. The ionization parameter $U$ 
($U$ = $Q /$ (4$\pi\,r^{2}\,c\,n\rm_{e}$)) is the number of available 
ionizing photons by hydrogen atom ($Q$ is the number of Lyman continuum photons 
emitted by the source per second, $r$ the distance of the emitting gas from 
the source, $n\rm_{e}$ the electron density of the cloud and $c$ the speed 
of light). Ferland \& Netzer (1983) showed that Seyfert 2s correspond to 
$U \sim$ 3\,10$^{-3}$ and Liners to $U \sim$ 3\,$10^{-4}$. 

Halpern \& Steiner (1983) suggested that dilution of the input continuum 
could be obtained if cold clouds with column density 
$N\rm_{H}$ = 10 $^{22}\,$cm$^{-2}$, 
typical of broad-line clouds in Seyfert 1 galaxies, were covering a fraction 
$f$ of the continuum as seen from the narrow-line region. The effect of 
covering is almost equal to a decrease in the ionization parameter $U$ by a 
factor (1$-f$). Liner spectra would be obtained for $f$ $\sim$ 0.90--0.98. 
Schultz \& Fritsch (1994) and Binette et al. (1996) have proposed very similar 
models to produce Liner spectra in which an average AGN continuum is 
distorted or ``filtered'' by matter-bounded clouds, hidden from view by 
obscuring material. An intervening ionized cloud with log $N\rm_{H}$ = 20 
would reduce the ionization parameter by a factor of $\sim$ 10. 
Binette et al. argue that, in these models, the predicted 
\ion{He}{ii}\,$\lambda$4686$/$H$\beta$ ratio is $<$~0.01, in 
agreement with the fact that no reliable detection of \ion{He}{ii} has been 
reported in Liners, while if $U$ is simply reduced, as proposed in the 
Halpern \& Steiner model, without altering the shape of the ionizing 
spectrum, the expected \rheii\ ratio is $\sim$ 0.15. In the case of 
NGC 5252, the \ion{He}{ii} line could not be firmly detected on the 
{\it HST} nucleus spectrum; nevertheless, an upper limit to the relative 
flux of this line to \hb\ can be estimated at $\sim$ 0.10, which does not 
allow us to decide between the two models. 

In NGC 5252, a Seyfert 2 and a Liner are simultaneously present. Both near-IR 
and X-ray observations reveal the presence of an obscured non-stellar nuclear 
source. This source is most probably responsible for the ionization of the 
Seyfert 2 nebulosity. We suggest that the Liner is also ionized by this 
nuclear source, attenuated by an intervening matter-bounded cloud hidden from 
view by the same material which obscures the nuclear source; the 
``filtering'' material would only partially cover the ionizing source, 
the Seyfert 2 clouds ``seeing'' directly this source without any intervening 
matter.\\ 
 

\begin{acknowledgements}

This research has made use of the NASA/IPAC extragalactic data\-base (NED), 
whi\-ch is opera\-ted by the Jet Pro\-pul\-sion Laboratory, Caltech, un\-der 
contract with the National Aeronautics and Space Administration. A.C. 
Gon\c{c}alves acknowledges support from the {\it Junta Nacional de 
Investiga\c{c}\~ao Cient\'{\i}\-fi\-ca e Tecnol\'ogica}, Portugal, during 
the course of this work (PhD. grant PRAXIS XXI/BD/5117/95).\\

\end{acknowledgements}


\end{document}